\documentclass[prl,aps,twocolumn,superscriptaddress,showpacs,preprintnumbers,
amsmath,amssymb,floatfix]{revtex4}

\def\gtwid{\mathrel{\raise.3ex\hbox{$>$\kern-.75em\lower1ex\hbox{$\sim$}}}}
\def\ltwid{\mathrel{\raise.3ex\hbox{$<$\kern-.75em\lower1ex\hbox{$\sim$}}}}

\usepackage{graphicx}
\usepackage{dcolumn}
\usepackage{bm}

\def\gev{GeV/c$^2$}
\def\eg{{{\em e.g.}}}
\def\perday{d$^{-1}$}
\def\perkg{kg$^{-1}$}
\def\iru{\perkg~\perday}
\def\etal{\it et al.\rm}
\long\def\symbolfootnote[#1]#2{\begingroup%
\def\thefootnote{\fnsymbol{footnote}}\footnote[#1]{#2}\endgroup}

\begin{document}

\title{First Results from the Cryogenic Dark Matter Search in the 
Soudan 
Underground Lab}


\author{D.S.~Akerib} \affiliation{Department of Physics, Case Western 
Reserve University, Cleveland, OH  44106, USA}
\author{J.~Alvaro-Dean} \affiliation{Department of Physics, 
University of 
California, Berkeley, CA 94720, USA}
\author{M.S.~Armel--Funkhouser} \affiliation{Department of Physics, 
University of 
California, Berkeley, CA 94720, USA}
\author{M.J.~Attisha} \affiliation{Department of Physics, Brown 
University, 
Providence, RI 02912, USA}
\author{L.~Baudis} \affiliation{Department of Physics, University of 
Florida, Gainesville, FL 32611, USA} \affiliation{Department of 
Physics, 
Stanford University, Stanford, CA 94305, USA}
\author{D.A.~Bauer} \affiliation{Fermi National Accelerator 
Laboratory, 
Batavia, IL 60510, USA} 
\author{J.~Beaty} \affiliation{School of Physics \& Astronomy, 
University 
of Minnesota, Minneapolis, MN 55455, USA}
\author{P.L.~Brink} \affiliation{Department of Physics, Stanford 
University, Stanford, CA 94305, USA}
\author{R.~Bunker} \affiliation{Department of Physics, University of 
California, Santa Barbara, CA 93106, USA}
\author{S.P.~Burke} \affiliation{Department of Physics, University of
California, Santa Barbara, CA 93106, USA}
\author{B.~Cabrera} \affiliation{Department of Physics, Stanford 
University, Stanford, CA 94305, USA}
\author{D.O.~Caldwell} \affiliation{Department of Physics, University 
of 
California, Santa Barbara, CA 93106, USA}
\author{D.~Callahan} \affiliation{Department of Physics, University 
of 
California, Santa Barbara, CA 93106, USA}
\author{J.P.~Castle} \affiliation{Department of Physics, Stanford 
University, Stanford, CA 94305, USA}
\author{C.L.~Chang} \affiliation{Department of Physics, Stanford 
University, Stanford, CA 94305, USA}
\author{R.~Choate} \affiliation{Fermi National Accelerator 
Laboratory, 
Batavia, IL 60510, USA}
\author{M.B.~Crisler} \affiliation{Fermi National Accelerator 
Laboratory, 
Batavia, IL 60510, USA}
\author{P.~Cushman} \affiliation{School of Physics \& Astronomy, 
University 
of Minnesota, Minneapolis, MN 55455, USA}
\author{R.~Dixon} \affiliation{Fermi National Accelerator Laboratory, 
Batavia, IL 60510, USA}
\author{M.R.~Dragowsky} \affiliation{Department of Physics, Case 
Western 
Reserve University, Cleveland, OH  44106, USA}
\author{D.D.~Driscoll} \affiliation{Department of Physics, Case 
Western 
Reserve University, Cleveland, OH  44106, USA}
\author{L.~Duong} \affiliation{School of Physics \& Astronomy, 
University 
of Minnesota, Minneapolis, MN 55455, USA}
\author{J.~Emes} \affiliation{Lawrence Berkeley National Laboratory, 
Berkeley, CA 94720, USA}
\author{R.~Ferril} \affiliation{Department of Physics, University of 
California, Santa Barbara, CA 93106, USA}
\author{J.~Filippini} \affiliation{Department of Physics, University 
of 
California, Berkeley, CA 94720, USA}
\author{R.J.~Gaitskell} \affiliation{Department of Physics, Brown 
University, Providence, RI 02912, USA}
\author{M.~Haldeman} \affiliation{Fermi National Accelerator 
Laboratory, 
Batavia, IL 60510, USA}
\author{D.~Hale} \affiliation{Department of Physics, University of 
California, Santa Barbara, CA 93106, USA}
\author{D.~Holmgren} \affiliation{Fermi National Accelerator 
Laboratory, 
Batavia, IL 60510, USA}
\author{M.E.~Huber} \affiliation{Department of Physics, University of 
Colorado at Denver, Denver, CO 80217, USA}
\author{B.~Johnson} \affiliation{Fermi National Accelerator 
Laboratory, 
Batavia, IL 60510, USA}
\author{W.~Johnson} \affiliation{Fermi National Accelerator 
Laboratory, 
Batavia, IL 60510, USA}
\author{S.~Kamat} \affiliation{Department of Physics, Case Western 
Reserve 
University, Cleveland, OH  44106, USA}
\author{M.~Kozlovsky} \affiliation{Fermi National Accelerator 
Laboratory, 
Batavia, IL 60510, USA}
\author{L.~Kula} \affiliation{Fermi National Accelerator Laboratory, 
Batavia, IL 60510, USA}
\author{S.~Kyre} \affiliation{Department of Physics, University of 
California, Santa Barbara, CA 93106, USA}
\author{B.~Lambin} \affiliation{Fermi National Accelerator 
Laboratory, 
Batavia, IL 60510, USA}
\author{A.~Lu} \affiliation{Department of Physics, University of 
California, Berkeley, CA 94720, USA}
\author{R.~Mahapatra} \affiliation{Department of Physics, University 
of 
California, Santa Barbara, CA 93106, USA}
\author{A.G.~Manalaysay} \affiliation{Department of Physics, Case 
Western 
Reserve University, Cleveland, OH  44106, USA}
\author{V.~Mandic} \affiliation{Department of Physics, University of 
California, Berkeley, CA 94720, USA}
\author{J.~May} \affiliation{Department of Physics, University of 
California, Santa Barbara, CA 93106, USA}
\author{R.~McDonald} \affiliation{Lawrence Berkeley National 
Laboratory, 
Berkeley, CA 94720, USA}
\author{B.~Merkel} \affiliation{Fermi National Accelerator 
Laboratory, 
Batavia, IL 60510, USA}
\author{P.~Meunier} \affiliation{Department of Physics, University of 
California, Berkeley, CA 94720, USA}
\author{N.~Mirabolfathi} \affiliation{Department of Physics, 
University of 
California, Berkeley, CA 94720, USA}
\author{S.~Morrison} \affiliation{Fermi National Accelerator 
Laboratory, 
Batavia, IL 60510, USA}
\author{H.~Nelson} \affiliation{Department of Physics, University of 
California, Santa Barbara, CA 93106, USA}
\author{R.~Nelson} \affiliation{Department of Physics, University of 
California, Santa Barbara, CA 93106, USA}
\author{L.~Novak} \affiliation{Department of Physics, Stanford 
University, 
Stanford, CA 94305, USA}
\author{R.W.~Ogburn} \affiliation{Department of Physics, Stanford 
University, Stanford, CA 94305, USA}
\author{S.~Orr} \affiliation{Fermi National Accelerator Laboratory, 
Batavia, IL 60510, USA}
\author{T.A.~Perera} \affiliation{Department of Physics, Case Western 
Reserve University, Cleveland, OH  44106, USA}
\author{M.C.~Perillo~Isaac} \affiliation{Department of Physics, 
University 
of California, Berkeley, CA 94720, USA}
\author{E.~Ramberg} \affiliation{Fermi National Accelerator 
Laboratory, 
Batavia, IL 60510, USA}
\author{W.~Rau} \affiliation{Department of Physics, University of 
California, Berkeley, CA 94720, USA}
\author{A.~Reisetter} \affiliation{School of Physics \& Astronomy, 
University of Minnesota, Minneapolis, MN 55455, USA}
\author{R.R.~Ross} \thanks{Deceased}
\affiliation{Lawrence Berkeley National 
Laboratory, 
Berkeley, CA 94720, USA} \affiliation{Department of Physics, 
University of 
California, Berkeley, CA 94720, USA}  
\author{T.~Saab} \affiliation{Department of Physics, Stanford 
University, 
Stanford, CA 94305, USA}
\author{B.~Sadoulet} \affiliation{Lawrence Berkeley National 
Laboratory, 
Berkeley, CA 94720, USA} \affiliation{Department of Physics, 
University of 
California, Berkeley, CA 94720, USA}
\author{J.~Sander} \affiliation{Department of Physics, University of 
California, Santa Barbara, CA 93106, USA}
\author{C.~Savage} \affiliation{Department of Physics, University of 
California, Santa Barbara, CA 93106, USA}
\author{R.L.~Schmitt} \affiliation{Fermi National Accelerator 
Laboratory, 
Batavia, IL 60510, USA}
\author{R.W.~Schnee} \affiliation{Department of Physics, Case Western 
Reserve University, Cleveland, OH  44106, USA}
\author{D.N.~Seitz} \affiliation{Department of Physics, University of 
California, Berkeley, CA 94720, USA}
\author{B.~Serfass} \affiliation{Department of Physics, University of 
California, Berkeley, CA 94720, USA}
\author{A.~Smith} \affiliation{Lawrence Berkeley National Laboratory, 
Berkeley, CA 94720, USA}
\author{G.~Smith} \affiliation{Department of Physics, University of 
California, Berkeley, CA 94720, USA}
\author{A.L.~Spadafora} \affiliation{Lawrence Berkeley National 
Laboratory, 
Berkeley, CA 94720, USA}
\author{K.~Sundqvist} \affiliation{Department of Physics, University 
of 
California, Berkeley, CA 94720, USA}
\author{J-P.F.~Thompson} \affiliation{Department of Physics, Brown 
University, Providence, RI 02912, USA}
\author{A.~Tomada} \affiliation{Department of Physics, Stanford 
University, 
Stanford, CA 94305, USA}
\author{G.~Wang} \affiliation{Department of Physics, Case Western 
Reserve 
University, Cleveland, OH  44106, USA}
\author{J.~Williams} \affiliation{Fermi National Accelerator 
Laboratory, 
Batavia, IL 60510, USA}
\author{S.~Yellin} \noaffiliation
\author{B.A.~Young} \affiliation{Department of Physics, Santa Clara 
University, Santa Clara, CA 95053, USA} 

\collaboration{CDMS Collaboration}

\noaffiliation

\date{\today}

\begin{abstract}
We report the first results from a search for weakly interacting 
massive particles (WIMPs) in the Cryogenic Dark Matter Search (CDMS) 
experiment at the Soudan Underground Laboratory. Four Ge and two Si 
detectors were operated for 52.6~live days, providing 19.4~kg-d of Ge 
net exposure after cuts for recoil energies between 10--100~keV.  
A blind analysis was performed using only calibration data to define 
the energy threshold and selection criteria for nuclear-recoil 
candidates.  
Using the standard dark-matter halo and nuclear-physics WIMP model,
these data set the world's lowest exclusion limits on 
the  coherent WIMP-nucleon scalar cross-section for all WIMP masses 
above 15~\gev, ruling out a significant range of neutralino 
supersymmetric models.
The minimum of this limit curve at the 90\% C.L. is $4 \times 
10^{-43}$ cm$^2$ at a WIMP mass of 60~\gev.
\end{abstract}


\maketitle

There is a compelling scientific case that nonluminous, nonbaryonic, 
weakly interacting massive particles 
(WIMPs)~\cite{lee,goodman,primack} may constitute most of the matter 
in the universe~\cite{bergstrom}. Supersymmetry provides a natural 
WIMP candidate in the form of the lightest superpartner, which must 
be stable if R-parity is 
conserved~\cite{jkg,ellis,kim,Baltz01,Bottino03}.  The WIMPs are expected 
to be in a roughly isothermal halo whose gravitational potential well 
contains the visible portion of our galaxy. These WIMPs would 
interact elastically with nuclei, generating a recoil energy of a few 
tens of~keV, at a rate smaller than $\sim$1~event~\iru 
~\cite{goodman,primack,jkg,ellis,kim,Baltz01,Bottino03,lewin}.

The Cryogenic Dark Matter Search (CDMS) collaboration is operating a 
new apparatus~\cite{r118prd} to search for WIMPs in the Soudan Underground 
Laboratory.  The CDMS Soudan experiment, also called CDMS~II, uses a 
set of Ge (each 250~g) and Si (each 100~g) ZIP (Z-dependent 
Ionization and Phonon) detectors~\cite{zips}, cooled to temperatures 
$< 50 $~mK and surrounded by substantial shielding deep underground
to
reduce backgrounds from 
radioactivity and cosmic-ray interactions. Simultaneous measurement 
of ionization and athermal phonon signals 
in the Ge and Si detectors allows excellent rejection of the 
remaining gamma and beta backgrounds. These background particles 
scatter off electrons in the detectors, while WIMPs (and neutrons) 
scatter off nuclei. The ZIP detectors allow discrimination between 
electron and nuclear recoils through two effects.  First, for a given 
energy, recoiling electrons are more ionizing than recoiling nuclei, 
resulting in a higher ratio of ionization to phonon signal, called 
``ionization yield."  Second, the athermal phonon signals due to 
nuclear recoils have longer rise times and occur later than those due 
to electron recoils.
For recoils within a few $\mu$m of a detector's surface (primarily 
from low-energy electrons), the charge collection is 
incomplete~\cite{shutt}, making discrimination based on ionization 
yield less effective. But these events can be effectively rejected by 
phonon timing cuts because they have, on the average, even faster 
phonon signals than those from bulk electron 
recoils~\cite{clarke,mandic}.  These effects are in qualitative 
agreement with our understanding of the complex phonon and 
semiconductor physics involved~\cite{cabrera}. 

The detectors are surrounded by an average of 0.5~cm of copper, 
22.5~cm of lead, and 50~cm of polyethylene, which reduce backgrounds 
from external photons and neutrons. 
A 5-cm-thick scintillator muon veto enclosing the shielding 
identifies charged particles (and some neutral particles) that pass 
through it.  An overburden of 780~m of rock, or 2090~meters water 
equivalent (mwe), reduces the surface muon flux by a factor of
$5 \times 10^{4}$.

All materials surrounding the detectors have been screened to 
minimize radioactive decays which could produce neutrons.  Neutrons 
resulting from radioactive decays outside the shield are moderated 
sufficiently to produce recoil energies below our detector 
threshold.  Neutrons produced in the shield by high-energy cosmic-ray 
muons are tagged by the veto scintillator with an efficiency 
$>$99\%.  The dominant unvetoed neutron background is expected to 
arise from neutrons produced by cosmic-ray muon interactions in the 
walls of the cavern.  
Events due to neutrons can be distinguished 
in part
from ones due to WIMPs 
because neutrons often scatter in more than one detector and interact 
at about the same rate in Si and Ge.  By contrast, WIMPs would not 
multiple-scatter, and coherent scalar WIMP interactions would occur 
$\sim$6$\times$ 
more often in Ge than in Si detectors.
 
We report here the analysis of the first CDMS Ge 
WIMP-search data taken 
at Soudan during the period October 11, 2003 through January 11, 
2004~\cite{r118prd}.  
After excluding time for calibrations, cryogen transfers, 
maintenance, and periods of increased noise,
we obtained 52.6~live days with the four Ge and two Si 
detectors of  ``Tower~1" (six close-stacked ZIP detectors labeled as 
Z1(Ge), Z2(Ge), Z3(Ge), Z4(Si), Z5(Ge) and Z6(Si) from top to 
bottom). Tower~1 was operated previously in an identical 
configuration at the Stanford Underground Facility (SUF), at a depth 
of 17~mwe~\cite{akerib03}. 

Energy calibrations were performed 
repeatedly during the run
using a $^{133}$Ba gamma source 
with distinctive,
penetrating
lines at 356~keV and 384~keV.  
The excellent agreement between 
data and Monte Carlo simulations  
and the observation of the 10.4~keV Ga line from neutron activation of 
Ge indicated
that the energy calibration was accurate 
and stable 
to within a few 
percent.
Observation of the predicted energy spectrum from a $^{252}$Cf source
confirmed the energy scale for nuclear recoils~\cite{r118prd}. 

The trigger rate on phonon signals for the WIMP-search data runs was 
$\sim$0.1~Hz, with a recoil-energy threshold of $\sim$2~keV 
($\sim$4~keV for Z1).  The muon veto signals were recorded in a time 
history buffer for each detector trigger.  The summed veto rate above 
threshold was $\sim$600~Hz, mainly due to ambient gammas. At this 
rate, we reject about 3\% of events with accidental veto coincidences 
in the 50~$\mu$s before a detector trigger.  Data-quality cuts 
reject  the $\sim$5\% of events that show any sign of higher 
pre-trigger noise or possible pile-up. 

\begin{figure}[tbp]
\begin{center}
\includegraphics[width=3.0in]{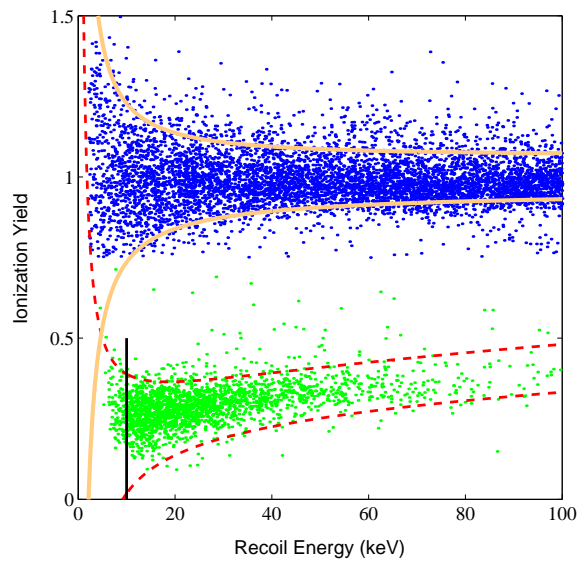}
\caption{\small Ionization yield versus recoil energy for calibration 
data with a $^{252}$Cf gamma and neutron source 
for detectors Z2, Z3 and Z5 in Tower~1 
showing the $\pm 2 \sigma$ gamma band (solid curves) and the 
$\pm 2 \sigma$ nuclear-recoil band (dashed curves)
for Z5, the detector with the worst noise of these three. 
Events with 
ionization yield $ < 0.75$ (grey) are shown only if they pass
the phonon-timing cuts. The vertical line is the 10~keV analysis 
threshold for these three detectors.}
\label{fig:Z235_Cf_tcutlowy}
\end{center}
\end{figure}

\begin{figure}[htbp]
\begin{center}
\includegraphics[width=3.0in]{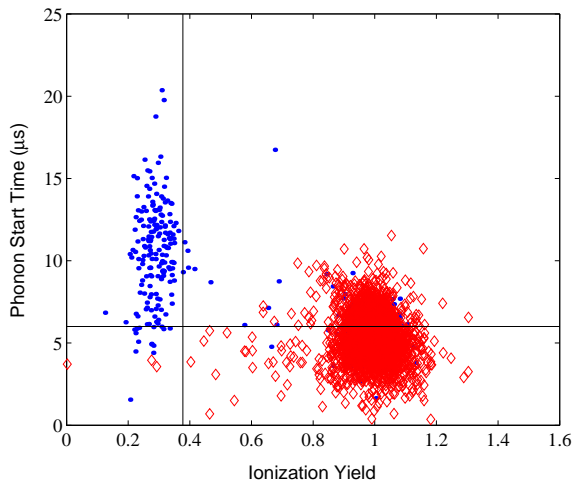}
\caption{\small Phonon start time versus ionization yield for 
$^{133}$Ba gamma-calibration events (diamonds) and 
$^{252}$Cf neutron-calibration events 
(dots) in the energy range 20--40~keV in detector Z5 in Tower~1.  
Lines indicate typical timing and ionization-yield cuts,
resulting in high nuclear-recoil efficiency and a low rate of 
misidentified surface events.}
\label{fig:Z5_t_y_Cfcals}
\end{center}
\end{figure}

\begin{figure}[htbp]
\begin{center}
\includegraphics[width=3in]{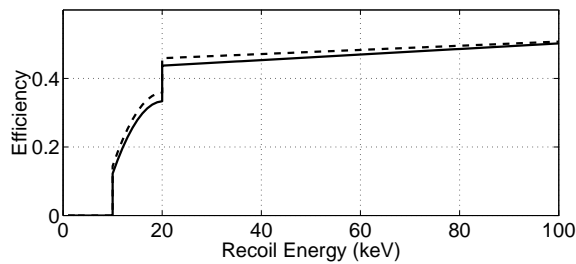}       
\caption{\small Efficiency of the combined cuts as a function of 
recoil energy, both for the blind analysis (solid) and for the 
second, non-blind analysis (dashed).  The step at 20~keV is due to 
Z1's 20-keV analysis threshold.}
\label{fig:cuteff}
\end{center}
\end{figure}

We performed a blind analysis, in which the nuclear-recoil region for 
the WIMP-search data was not inspected until all cuts and analysis 
thresholds were defined using {\it in situ} gamma and neutron 
calibrations (see Fig.\ \ref{fig:Z235_Cf_tcutlowy}). A combination of 
ionization-yield and phonon-timing cuts rejects virtually all 
calibration electron recoils while accepting most of the nuclear 
recoils.  The phonon timing cuts are based on both 
the phonon rise time and the phonon start time relative to the 
ionization signal
(see Fig.~\ref{fig:Z5_t_y_Cfcals}).  
We required recoil energy between 10--100~keV for all Ge 
detectors except Z1, whose larger noise 
required an 
analysis threshold of 20~keV in order to ensure
comparable rejection. 
We rejected events with some ionization in a detector's annular ``guard'' 
electrode, which covers 15\% of the detector's volume.
Figure~\ref{fig:cuteff} shows a 
conservative estimate of the combined efficiency of all cuts on a 
WIMP signal.  The cuts yield a spectrum-averaged effective exposure 
of 19.4~kg-days between 10--100~keV for a 60~\gev~WIMP.

\begin{table}[tdp]
\caption{Unvetoed 
gamma and surface-electron-recoil rates 
between 15--45~keV in Tower~1 at Soudan.}
\begin{center}
\begin{tabular}{|l|rcr|r|c|c|}
\hline
    & \multicolumn{4}{|c|}{Gammas [\#/kg/day]} & 
      \multicolumn{2}{|c|}{Surface [\#/day]} \\
 \hline
ZIP    & \multicolumn{3}{|c|}{(total)} & \multicolumn{1}{|c|}{(singles)} & 
(total)        & (singles)     \\
\hline   
Z1(Ge) & 85.6 & $\pm$ &3.4 &  37.6 $\pm$ 2.3 &  1.56 $\pm$ 0.23 & 
0.90 $\pm$ 0.18 \\
Z2(Ge) & 79.4 & $\pm$ &3.1 &  19.7 $\pm$ 1.6 &  1.05 $\pm$ 0.18 & 
0.18 $\pm$ 0.08 \\
Z3(Ge) & 89.3 & $\pm$ &3.3 &  19.9 $\pm$ 1.5 &  1.11 $\pm$ 0.18 & 
0.15 $\pm$ 0.07 \\
Z5(Ge) & 105.7& $\pm$ &3.6 & 35.9 $\pm$ 2.1 &  1.82 $\pm$ 0.24 & 
0.65 $\pm$ 0.14 \\
\hline
\end{tabular}
\end{center}
\label{rates}
\end{table}

Table~\ref{rates} lists the observed rates of 
unvetoed
events in the WIMP-search data, with ionization yield either in the 
$\pm2\sigma$ gamma band (``gammas") or 
below this band (mostly surface electron recoils).  Analysis shows 
that about half the surface electron recoils with interactions in 
only a single detector (``singles'') were due to beta decays of 
contaminants on surfaces, while the other half were from 
gamma rays.
Gamma rates are $\sim$50\% higher at Soudan than they 
were at SUF, consistent with the higher Rn levels 
at Soudan and
the absence of a 1-cm-thick ancient-Pb liner which surrounded the 
detectors at SUF. Total surface-event rates at Soudan are also somewhat higher 
than at SUF, consistent with the increased
component due to gammas.

We computed the number of electron-recoil events expected to be 
misidentified as nuclear recoils in the WIMP-search data
based on the  $^{133}$Ba calibration sets used to determine the 
timing cuts.  
Factoring in systematic errors, we estimated 
$0.4 \pm 0.3$ misidentified events in Z1 and a total of
$0.3 \pm 0.2$ 
in the other Ge detectors.
As a 
check, we applied the same cuts to a different set of $^{133}$Ba calibrations,
containing 1.5 
times as many surface events as in the WIMP-search data. 
One event (at 50~keV in Z1) passed all cuts, in agreement with the 
previous estimate.

Monte Carlo simulations predict $0.05\pm0.02$ neutrons (mostly unvetoed)
produced from muon interactions outside the shielding, 
including uncertainties on the neutron production rate.
The simulations predict $\sim$1.9 (veto-coincident) neutrons 
produced inside the 
shielding for the WIMP-search data.  
No veto-coincident nuclear-recoil 
candidates were observed in the WIMP-search data.

\begin{figure}[tbp]
\begin{center}
\includegraphics[width=3.0in]{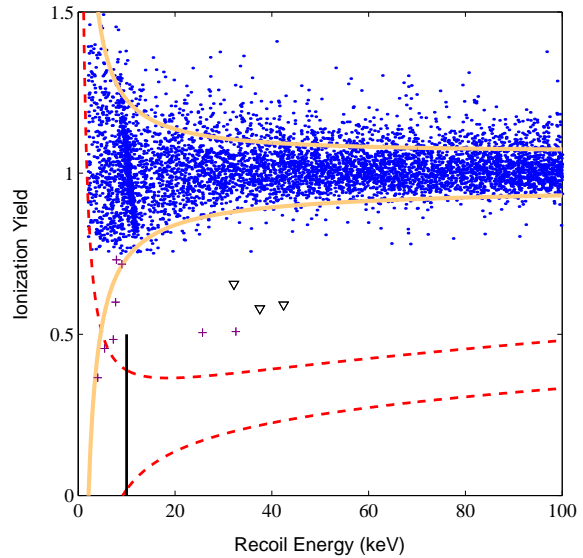}
\caption{\small Ionization yield versus recoil energy for WIMP-search 
data from Z2 (triangle), Z3, and Z5 ($+$) in Tower~1, 
using the same yield-dependent cuts and showing the same curves as in 
Fig.\ \ref{fig:Z235_Cf_tcutlowy}.
Above an ionization 
yield of 0.75, 
the events 
from all three detectors are drawn as identical points in order to 
show the 10.4 keV Ga line from neutron activation of Ge.}  
\label{fig:Z235_after_tcuts}
\end{center}
\end{figure}

\begin{figure}[htbp]
\begin{center}
\includegraphics[width=3.0in]{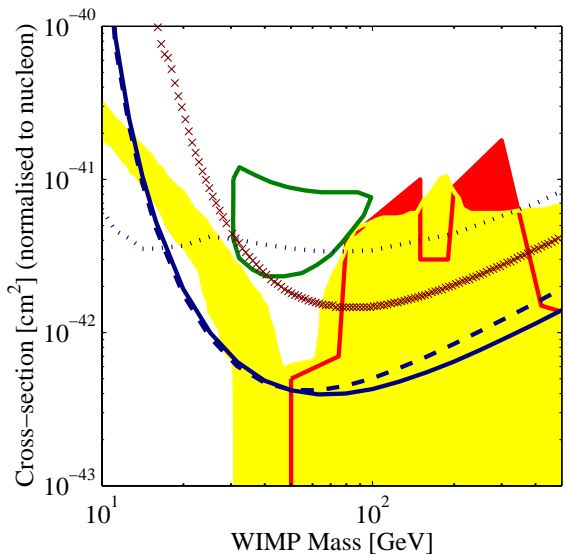}
\caption{\small New limit  on the WIMP-nucleon scalar cross section 
from CDMS~II at Soudan with no candidate events in 19.4 kg-d 
effective Ge exposure (solid curve). 
Parameter space above the curve is excluded at the 90\%
C.L.
These limits constrain
supersymmetry models, for example~\cite{Baltz01} (dark grey) 
and~\cite{Bottino03} (light grey).
The DAMA (1-4) 3$\sigma$ signal region~\cite{DAMA2000} is shown as a closed 
contour. 
Also shown are limits from CDMS at SUF~\cite{akerib03} (dots), 
EDELWEISS~\cite{edel0203} ($\times$'s), and the second, non-blind 
analysis of CDMS~II at Soudan with 1 nuclear-recoil candidate event 
(dashes).  All curves~\cite{limitplotter}
are normalized following~\cite{lewin}, using 
the Helm spin-independent form-factor, $A^2$ scaling, WIMP 
characteristic velocity $v_0 = 220 {\rm \; km \; s^{-1}}$, mean Earth 
velocity $v_E = 232 {\rm \; km \; s^{-1}}$, and $\rho = 0.3 {\rm \; 
GeV \; c^{-2} \; cm^{-3}}$. }
\label{fig:CDMS_prelimit_2004}
\end{center}
\end{figure}

This blind analysis of the first Soudan CDMS~II WIMP-search data set 
revealed no nuclear-recoil events in 52.6 kg-d raw exposure in our Ge 
detectors.  Figure~\ref{fig:Z235_after_tcuts} displays the ionization 
yield of WIMP-search events in Z2, Z3, and Z5 which passed the
same cuts applied to calibration data in Fig.\ \ref{fig:Z235_Cf_tcutlowy}.  
As shown in 
Fig.\ \ref{fig:CDMS_prelimit_2004}, these data
together with corresponding data 
for Z1 set an upper limit on the WIMP-nucleon cross-section of $4 
\times 10^{-43}$~cm$^2$ at the 90\%~C.L.  at a WIMP mass of 
60~\gev~for  
coherent scalar interactions
and a standard WIMP halo. 

After unblinding the nuclear-recoil region, we found that our 
pulse-fitting algorithm designed to handle saturated pulses had been 
inadvertently used to analyze 
most of the unsaturated pulses in the WIMP-search data. This 
algorithm gives slightly worse energy resolution than 
the intended algorithm. The limit in 
Fig.~\ref{fig:CDMS_prelimit_2004} based on the blind analysis 
(solid line) correctly accounts for this effect. We have also 
performed a second, non-blind analysis, 
using the intended pulse-fitting algorithm and the same blind cuts,
resulting in a 5\% higher WIMP detection efficiency. This analysis 
resulted 
in
one nuclear-recoil candidate (at 64~keV in Z5), consistent with the expected
surface-event misidentification quoted above.
Figure~\ref{fig:CDMS_prelimit_2004} 
includes the optimum interval \cite{yellin} limit based on this 
second unbiased, but non-blind, analysis (dashed line).

At 60~\gev, these limits are a factor of four below the best previous 
limits set by 
EDELWEISS~\cite{edel0203}, and a factor of eight better than our 
limit with the 
same Tower~1 at SUF~\cite{akerib03}.  These data confirm that events 
detected by CDMS at SUF and those detected by EDELWEISS were not a 
WIMP signal. 
Under the assumptions of a standard halo model,
our new limits are clearly incompatible with
the DAMA (1-4) signal region~\cite{DAMA2000}
if it is due to coherent scalar WIMP interactions (for DAMA regions
under other assumptions see~\cite{DAMA2003}).
Our new limits significantly constrain supersymmetric models under  
some theoretical frameworks that place weak constraints on 
symmetry-breaking parameters (\eg\ \cite{kim,Baltz01,Bottino03}).

This work is supported by the National Science Foundation under Grant 
No.\ AST-9978911, by the Department of Energy under contracts 
DE-AC03-76SF00098, DE-FG03-90ER40569, DE-FG03-91ER40618, and by 
Fermilab, operated by the Universities Research Association, Inc., 
under Contract No.\ DE-AC02-76CH03000 with the Department of Energy.  
The ZIP detectors were fabricated in the Stanford Nanofabrication 
Facility operated under NSF. 
We are grateful to the Minnesota Department of Natural Resources 
and the staff of the Soudan Underground Laboratory for their 
assistance.

\gdef\journal#1, #2, #3, #4#5#6#7{ #1~{\bf #2}, #3 (#4#5#6#7)} 

\def\arp{\journal Ann.\ Rev.\ Nucl.\ Part.\ Sci., }
\def\apl{\journal Appl.\ Phys.\ Lett., }
\def\apj{\journal Astrophys.\ J., }
\def\app{\journal Astropart.\ Phys., }
\def\baas{\journal Bull.\ Am.\ Astron.\ Soc., }
\def\ejpc{\journal Eur.\ J.\ Phys.\ C., }
\def\jltp{\journal J.\ Low\ Temp.\ Phys., }
\def\jhep{\journal J.\ High Energy Phys., }
\def\nature{\journal Nature, }
\def\nc{\journal Nuovo Cimento, }
\def\nima{\journal Nucl.\ Instr.\ Meth.\ A, }
\def\np{\journal Nucl.\ Phys., }
\def\npps{\journal Nucl.\ Phys.\ (Proc.\ Suppl.), }
\def\pl{\journal Phys.\ Lett., }
\def\prep{\journal Phys.\ Rep., }
\def\pr{\journal Phys.\ Rev., }
\def\prc{\journal Phys.\ Rev.\ C, }
\def\prd{\journal Phys.\ Rev.\ D, }
\def\prl{\journal Phys.\ Rev.\ Lett., }
\def\rnc{\journal Riv.\ Nuovo\ Cim., }
\def\rsi{\journal Rev. Sci. Instr., }
\def\rpp{\journal Rep.\ Prog.\ Phys., }
\def\sjnp{\journal Sov.\ J.\ Nucl.\ Phys., }
\def\solarphys{\journal Solar Phys., }
\def\jetp{\journal J.\ Exp.\ Theor.\ Phys., }

\end{document}